# Lost Again in Shibuya:
# Exploration and Awareness in a Labyrinth[1]

**Shin'ichi Konomi**
Center for Spatial Information Science
The University of Tokyo
5-1-5, Kashiwanoha, Kashiwa, Chiba 277-8568, JAPAN
konomi@csis.u-tokyo.ac.jp

## ABSTRACT
Existing digital technologies in urban settings tend to focus narrowly on concerns around wayfinding, safety, and consumption. In this paper, we examine pedestrian experiences based on the data collected through field observations as well as intensive interviews with nine pedestrians in the Shibuya area of Tokyo, and suggest an alternative approach to blending technologies and urban activities. Our focus is on social and cognitive aspects of pedestrians who get lost and explore a labyrinth of sidewalks. We use the data to discuss the activities that are often ignored or inadequately supported by existing systems.

## INTRODUCTION
When people talk about digital technologies in public spaces (e.g., sidewalks, train stations, coffee shops, etc.), they may refer to ATM machines, kiosk terminals, CCTV cameras, RFID tickets, public displays showing advertisements, or mobile phone-based pedestrian navigation systems. These technologies generally increase efficiency and transparency, and arguably transform public spaces into a busier site of consumption.

Our work underlines the importance of a broader view of public-space technologies by closely examining our local urban environment: the Shibuya area of Tokyo. The area is best known as a center stage of colorful youth cultures and Shibuya's scramble crossing (see Figure 1) often appears in TV shows and movies, including the American film *Lost in Translation*.

The scramble crossing as well as the nearby sidewalks and the pedestrian shopping street called *Center Gai* is the most packed with advertisement messages from large multimedia displays, billboards, sidewalk speakers, moving trucks, solicitors from restaurants and karaoke rooms, *"Tissue Kubari"* people handing out free ad pocket tissues, etc. Shibuya can be described as "a place for browsing" [15] in which consumers interact with information, things (e.g., sales items), and people.

Shibuya's complex spatial layout is a labyrinth [7] (p.61) rather than a grid. It creates a situation for pedestrians to get lost, wander and explore. Once people walk into the narrow, visually-limited pedestrian streets, they are forced to rely on memory and other clues in order to master the area. Such 'non-geometric' spaces could be understood as 'cinetic space' [7], in which "the experience of moving through counts more than the destination."

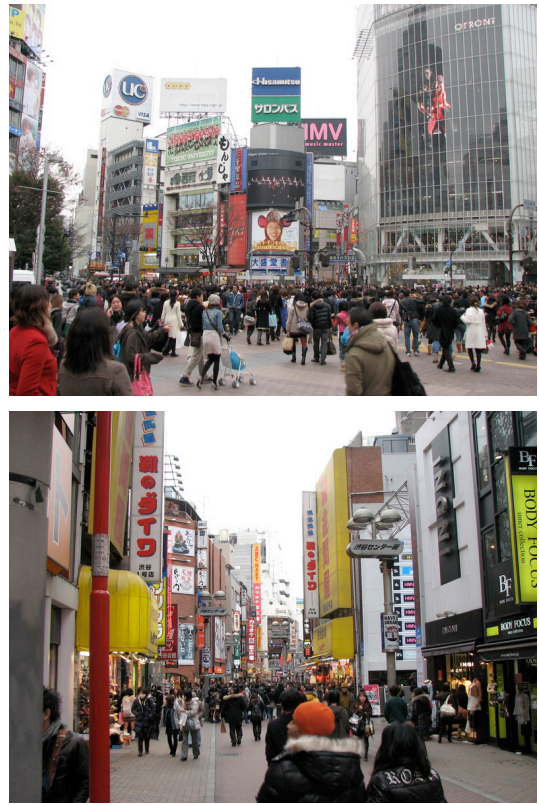

**Figure 1. Shibuya's scramble crossing as viewed from the train station (top), and *Center Gai* (bottom).**

Given the diversity of people and their activities in cities, many existing technologies in public spaces tend to focus narrowly on consumers who must travel from point A to B as efficiently and safely as possible without getting bored (and spend as much money as possible). Researchers recently looked into different types of users in London, UK [12] and Minneapolis, USA [10], who actively *collect, keep* and *share* information. Although these types of users could exist in many places in the world, we must take the impact of geographical context and cultural dimensions [4] seriously. We then should avoid blind adoption of 'universal' user and usage models, and turn to local specificities, which we can seek in ethnographic data.

---

[1] Most part of this paper was written in 2008. Minor revision was made in 2014.

**METHOD**

To discover how pedestrians think and act, we used techniques for gathering qualitative data through observations and interviews. Within the scope of our study, existing quantitative data (e.g., census data) lack richness of details even though they provide an important macro-level context for our analysis.

Inspired by the Urban Probes methodology [13], we began with field observations at the scramble crossing and a coffee shop facing a busy sidewalk near the crossing as well as a similarly busy space in Shinjuku (JR Shinjuku East Exit), which is also in Tokyo. These observations took place on sunny Saturday afternoon (Shibuya) and evening (Shinjuku), in the first week of July, 2007.

As a next step of study, Urban Probes encourage direct interventions. We however felt that our understanding about Shibuya was still quite superficial in terms of the diversity of subjective experiences. Therefore, we adopted a variation of intensive interviewing, a user study technique for gathering qualitative data by asking users open-ended questions about their work, background, and ideas [1].

We used the results of the observations to loosely structure the interviews with six graduate students and three university staff members who work in a nearby building (about 2km away from the scramble crossing). The interviewees include five locals who live in the Tokyo area for more than 20 years, and the others live, work or study in the area for at least two years. The total time we spent for each interviewee was between one to four hours. We coded and analyzed over 15 hours of interview data (i.e., audio recordings and annotated maps) considering existing theories (e.g., [6][7][9]) as well as the insights gained from existing cultural studies [11] and media discourses about the area.

It is also important to mention certain limitations of our data. Without interventions [13], our data can tell primarily about urban life in authentic environments. Therefore, we may run the risk of being overly conservative in forming ideas about how things could be. Nevertheless, we believe our data can inspire unique, unexplored ideas, as our geographical and cultural context is not thoroughly considered in existing mobile HCI research.

Our interviewees apparently do not represent *all* kinds of people in Shibuya. In particular, one may find many younger people in the area: teenagers and young adults in their early twenties. In our case, more than half interviewees use Shibuya since the time when they were kids, which should increase the diversity of experiences embodied in the interview data.

**FINDINGS**

Several themes emerged from our observation and interview data. Overall, we learned that people's activities are much more diverse than we first imagined, and that some social and cognitive experiences are inseparable from the area's unique physical patterns.

**Image**

It is not only the built environment that makes the image of Shibuya. Not surprisingly, people and their fashion styles, shops, as well as public and personal events play equally important roles in making the image. As Shibuya is well-known nationally, the mass media often talk about the area, influencing the perceptions of many people. Moreover, dynamic elements such as crowd movement patterns at the scramble crossing and even the images shown on large multimedia displays can remain vividly in one's memory.

The emerged image is collectively a mixture of fun, busyness and risks that makes tolerance an essential skill. Crowding and noisiness are the two major factors that sometimes keep people away from the area. There are also moderate concerns about extortion and malicious street solicitation in places like *Center Gai*, which may not keep people away but somewhat increase wariness towards strangers. Despite some negative perceptions, the area offers a vast array of things that keep many people coming.

**Use Patterns**

People from various age groups go to Shibuya for shopping, entertainment (e.g., movies) and dining. Young adults in their early twenties, in particular, often use the area for social activities such as partying. People choose the area over other similar areas because it is conveniently located in terms of public transportation, it offers more things to do than other areas, and their friends want to go there.

Most of our interviewees usually know what they want to do in Shibuya (and sometimes where they want to go as well) before they arrive in the area. One interviewee hangs out in coffee shops and other spaces much more frequently than others. An interesting contrast is drawn between their behaviors and their perceptions of others' behaviors: they think there are many people who just hang out in the area. Sometimes people go online and search places of interest before making a trip. Other times they wander about and look for places *in situ* using varieties of clues: photos and price information outside restaurants, queues of people waiting in front of shops, and information provided by street solicitors, just to name a few.

The diversity of home ranges (in Shibuya) we observed in the interview data indicates that usage patterns of Shibuya are different person by person. As individuals develop their patterns with familiar places and paths, it may become easier to get around, however, more difficult (or less motivated) to discover new things that do not intersect with established patterns.

**Lost "as Usual"**
It's really easy to get lost in Shibuya. Everyone, except for a local who frequented the area with parents for a long time, got lost in the area. Two locals say that they can still get lost even though they have been to the area hundreds of times. Our interviewees' comments suggest that even 'experts' get around primarily based on *route knowledge* rather than *survey knowledge* [3]:

*It's just a kind of feeling that I'm around here. I don't think like "I'm on street so-and-so and if I make a turn here I'll be on street so-and-so," you know. [I] just kind of figure out like, "oh, this scenery is around here, and it may be good to go this way" or...[2]*

When people get lost in Shibuya, they just walk around for a while till they find something familiar. They also use other strategies such as asking strangers, reading maps on sidewalks and mobile phones, and asking friends on the phone; however, these seem to be less frequently used than walking around. Two interviewees commented on the positive aspects of getting lost:

*Should I call it getting lost? I sometimes enjoy getting lost ... when I come across a familiar place, it's like "I made it!" and I really like that experience.*

*It's kind of, you know, when I walk around, I might discover something new.*

Shibuya's density of people, things and information seems to support these playful and explorative attitudes. Also, most of our interviewees think that Shibuya's physical structure with clear edges, unique landmarks, and a valley (if you walk downward, you get to the crossing) provides a safety net for preventing getting lost too much.

**Interacting with Strangers and Friends**
The scramble crossing as well as *Center Gai* is notoriously restless with people who are eager to talk to strangers: solicitors from restaurants and karaoke rooms, model and talent scouts, boys interested in getting to know stranger girls, and so on. When it's difficult to assess strangers' trustworthiness based on perceivable clues, people may decide to ignore them by walking faster or maintaining the same speed, not looking at them, or saying 'sorry.'

People may find it difficult to talk to strangers in such an environment. Suppose one gets lost and desperately needs immediate help from someone to be in time for a meeting. Whom would s/he ask for directions? Our interview data indicates that different people trust (and believe that they are trusted by) different types of strangers.

As a few of our interviewees suggested, it would be easier to meaningfully interact with like-minded strangers in places such as clubs, shops, and movie theaters. People seem to recognize familiar strangers in these and other similar places more than crowded places such as crossings, sidewalks, and train stations. Our interviewees do come across their friends in Shibuya, however, not very frequently. They think that they might have missed their friends who were actually in proximity.

**DISCUSSION**
Technology design for public spaces is a controversial issue. In this section, we discuss how our findings lead us to a unique approach to designing computing systems for pedestrians. Although "getting lost" and "dealing with strangers" are often considered as problematic situations that must be addressed, we argue that these situations can provide essential qualities in urban pedestrian experiences and technologies can be designed to support such qualities.

**Urban Learning**
Urban sociologist Lyn H. Lofland discusses positive aspects of urban environments with respect to the six 'uses' of "public realm" [9], i.e., (1) an environment for learning, (2) respites and refreshments, (3) a community center, (4) the 'practice' of politics, (5) the enactment of social arrangements and social conflict, and (6) the creation of cosmopolitans. Being aware that some social activities in non-Western public spaces could not be easily understood by only using the Western notion of a public sphere (cf. [6]), some of the unique factors we found in Shibuya seem to profoundly impact urban learning, the process of acquiring the requisite knowledge and skills for acting in a world of strangers [8].

Our findings suggest that mild, occasional disorientation can lead to enjoyable learning experiences for acquiring new knowledge and skills if it happens at the right time and the right place. Getting lost in "labyrinth cities" such as Shibuya could potentially be an experience that serves as a catalyst for subverting and transcending suboptimal, old activity patterns in dynamic urban spaces. Playful and explorative attitudes can emerge around such experiences.

Similarly, restless spaces such as Shibuya's scramble crossing and *Center Gai* can teach tolerance and create cosmopolitans, as Lofland [9] (p.243) puts it:

*Cleaned-up, tidy, purified, Disneyland cities (or sections of cities) where nothing shocks, nothing disgusts, nothing is even slightly feared may be pleasant sites for family outings or corporate gatherings, but their public places will not help to create cosmopolitans.*

Existing systems such as mobile phone-based pedestrian navigation applications are useful for 'visitors' who come from elsewhere and go elsewhere. Our Shibuya study suggests alternative applications intended for 'dwellers' who frequent it, find meanings in the experiences of walking within it, and learn from the experiences.

---

[2] Comments are originally in Japanese. They were translated into English by the author.

**Association and Awareness**
People's route-based knowledge creates powerful yet loosely constructed image of Shibuya. Such loosely arranged but not well-structured cognitive maps or *cognitive associational maps* [6] can facilitate informal non-geometrical association between spaces. Popular map-based user interfaces could complement Shibuya pedestrians' knowledge; however, they may not necessarily amplify (or be integrated with) their knowledge. To improve the quality of walking experiences, an alternative user interface may enhance exploration and informal association through awareness support. For example, 'egocentric' visualization of personal, social and historical aspects of a user's nearby space could augment her awareness about the space in various ways.

Shibuya is already overloaded with stimulus. Digital technologies such as public displays and mobile phones seem to merely saturate the space with more information and stimulus. However, digital technologies can also filter information and social interaction opportunities. We believe that tools for supporting awareness can be integrated with information filtering and social matching mechanisms that reduce "urban overload."

**CONCLUSION**
We took a close look at a particular urban setting and discussed pedestrian experiences. Of course, every urban setting is unique, and we can learn from other cities as much as we learned from the specificities of Shibuya. This implies that our understanding on the diversity of pedestrian experiences is still limited. For example, some Asian cities are similarly complex as in Shibuya; however, anecdotally, pedestrian behaviors seem somewhat different. Our data nevertheless informed us of the unique cognitive and social aspects that seem to be neglected by existing computing technologies in public spaces. We hope our work will inspire design of novel pedestrian applications that support various human activities besides wayfinding, consumption and advertisement. Walking and wandering may allow people to explore information, ideas and social interactivity in a more spontaneous, embodied, and improvisational manner than in disembodied media spaces.

We plan to conduct an intervention-based study based on the result of this research. In parallel, we are designing an integrated tool for enhancing pedestrian experiences in "labyrinth cities" such as Shibuya. Our proof of concept prototype uses active RFID technology to digitally (and socially) enhance pedestrians' egocentric views of the spaces they move through. It senses nearby people and anonymously visualizes their mobile blog contents (about places, things, and people they encounter as well as their general thoughts and feelings) based on geometric interpersonal distances. We are also exploring a way to incorporate non-geometric distances based on social networks and user preferences.